Budapest, Hungary, 17-19 September 2007# Micro-Hotplates for Thermal Characterisation of Structural Materials of MEMS

P. Fürjes[1], P. Csíkvári[2], I. Bársony[1] and Cs. Dücső[1]
[1] Research Inst. for Technical Physics and Materials Science – MFA, P.O.Box. 49, H-1525 Budapest
[2] Budapest University of Technology and Economics, Department of Atomic Physics
Budafoki str. 8., H-1111 Budapest, Hungary
E-mail: furjes@mfa.kfki.hu**Accurate** knowledge of mechanical and thermal properties of structural materials used in MEMS is essential for optimum geometric and functional design. The extraction of precise physical properties is rather complicated due to the size effects, the complexity of the structures and the variations of formation processes. This work is intended to determine the thermal properties of silicon-nitride and diamond layers applied in thermal sensor structures by analyzing thermal responses of a multilayer micro-heater structure.

*Keywords: micro-heater, material properties, thermal analysis*## I. INTRODUCTION

Micro-hotplates are basic components of sensors and lab on a chip devices, e.g. as sensors of calorimetric principle [1, 2], or heaters in chemical micro-reactors. The most frequently used structural materials are silicon-nitride, non-stoichiometric silicon-nitride, silicon-oxinitride, silicon-dioxide and multilayered combination of these materials. Beyond the frequently applied structural materials, protective coating of MEMS elements used in harsh environment is essential for their reliable operation. The best candidates for such application are diamond, diamond-like-carbon (DLC) or SiC because of their superior chemical and abrasion resistance in aggressive chemicals.

In the micro-hotplate design the most important parameters to be considered are the thermal conductivity, the thermal capacitance and the residual stress in the applied layers in order to select the optimum functionality of the device. While appropriate data are available for the widely used materials ($SiO_2$, $Si_3N_4$) this is far not being the case for the non-stoichiometric materials or deposited diamond and DLC layers. Their properties are process dependent, i.e. both their composition and structure are determined by the given individual process. Therefore, relatively simple methods for determination of thermo-mechanical properties are essential in functional design.

In this work we describe alternative techniques for the formation of micro-filaments encapsulated in CVD non-stoichiometric silicon-nitride or multi-crystalline CVD diamond layers and present how the thermal properties of the structural materials can be determined by measuring static and dynamic temperature responses of the micro-hotplate.

## II. TEST STRUCTURES AND PROCESSES

Operation of micro-hotplates at elevated temperature requires adequate thermal isolation of the heated surface in order to reduce the power dissipation to a reasonable level. In our approach two arm suspended hotplate structures were formed from reduced stress Si-rich silicon-nitride and MW-PE-CVD (Microwave Plasma Enhanced Chemical Vapour Deposition) microcrystalline diamond layers with embedded platinum or poly-crystalline silicon filaments, respectively. The micro hotplates ($100 \times 100 \times 1 \mu m^3$) are suspended across a 60-80μm deep cavity formed by selective dissolution of an electrochemically etched porous silicon sacrificial layer. For enhanced mechanical stability of the suspended filament structure a slim blade-like silicon supporting pillar can also be formed under the centre axis of the plate by exploiting the reduced etching rate of Si when the two isotropic etching fronts approach each other to a distance of the actual depletion layer thickness. The ~0.3-0.5μm thick mechanical support provides excellent thermal insulation due to the deep electronic depletion. (Fig.1.)

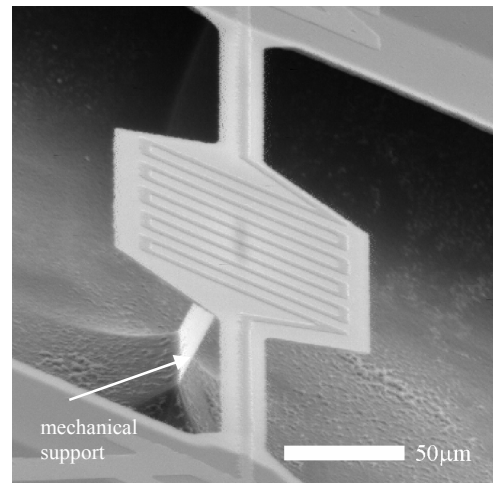

Fig. 1. SEM view of the silicon-nitride based micro-filament heater as basic element of a calorimetric type sensor.

©EDA Publishing/THERMINIC 2007 -page- ISBN: 978-2-35500-002-7

In Version 1 the Pt heater filaments are embedded in silicon rich silicon-nitride double-layer. The reduced stress layer is deposited by LPCVD process at 830°C from a $SiH_2Cl_2:NH_3=5:1$ gas mixture. The detailed process is described in [3, 4].

The diamond coated filament structure (Version 2) also utilizes the porous silicon micromachining [4] technique in combination with a three-step multilayer deposition (diamond/poly-crystalline silicon/diamond), doping and appropriate patterning. The material of the filament is preferably poly-crystalline silicon in order to easy adaptation of the Selective Area Deposition (SAD) technique used for patterning [5]. The main steps of the process sequence are presented by Fig. 2.

Diamond layers were formed by MW-PE-CVD technique, in a conventional quartz bell-jar reactor, applying BEN (Bias Enhanced Nucleation). The source gases were methane and hydrogen. The nucleation was performed under a pressure of 25mbar, while the substrate temperature was 750°C. To achieve selective area deposition, the bias enhanced nucleation parameters were optimized, i.e.: the methane concentration, the bias and the biasing time. Appropriate setting of these parameters is required to minimize the milling of the masking layer ($SiO_2$ and $Si_3N_4$) since the reactive, accelerated species during BEN can sputter the surface, otherwise the selectivity of the deposition is lost. The optimal parameters for the biasing were the follows: 15 min under -200V bias and the methane concentration was 10%-12%, which was necessary to assure a sufficiently fast nucleation.

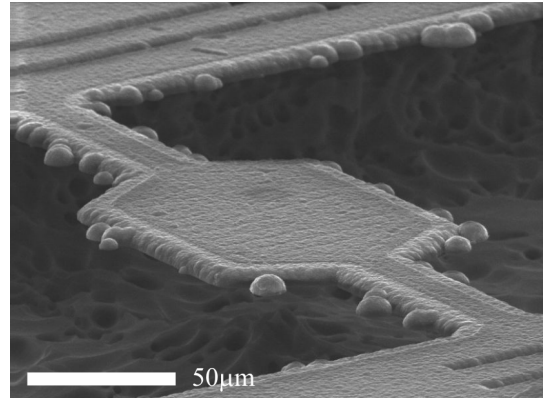

Fig. 3. SEM view of the diamond coated micro-filament heater.

After the nucleation, a 3 hour growing cycle was applied at 40 mbar pressure and 1200 W microwave power. To avoid the thermal stress induced cracking of the masking $Si_3N_4$ layer the substrate temperature was the same as in the nucleation process i.e. 750°C. As the masking layer can be sputtered even during the growing procedure, the growth rate of diamond must be high enough to prevent the nitride mask and to achieve a relatively fast and pinhole free diamond layer formation. For this reason, the methane concentration was kept at 1%-2%, which results in a good quality, polycrystalline, pinhole free diamond.

In the second stage of the process, the selective area deposition was repeated; however, modified conditions were applied. Since a very thin poly-silicon filament and contact wires were formed on the diamond coated area, their sputtering had also to be avoided. For that reason, in a very high growing rate without any biasing was applied in order to have the poly-silicon wires (ca. 700nm high step) hemmed in by coalescent diamond layer. To reach this aim, at 750°C temperature, with 1400 W microwave power, 50 mbar pressure and relatively high, 2%-3% methane concentration was applied. After 10-12 hours growing the layer closed round the poly-Si meanders.

### III. RESULTS AND DISCUSSION

*Extraction of Physical Parameters*

The thermal properties of the applied structural materials were determined by detailed thermal analysis of sensor structures. The combination of different thermal analysing techniques and accurate model calculations of the structures is applicable for extraction of the physical parameters. [6] Utilising the measured results of the static and dynamic thermal analysis for the simulation step, the real physical parameters of the applied materials can be practically estimated by scanning an adequate parameter range during the model calculation. Getting accurate results the precise model construction is absolute necessary, particularly the real dimensions.

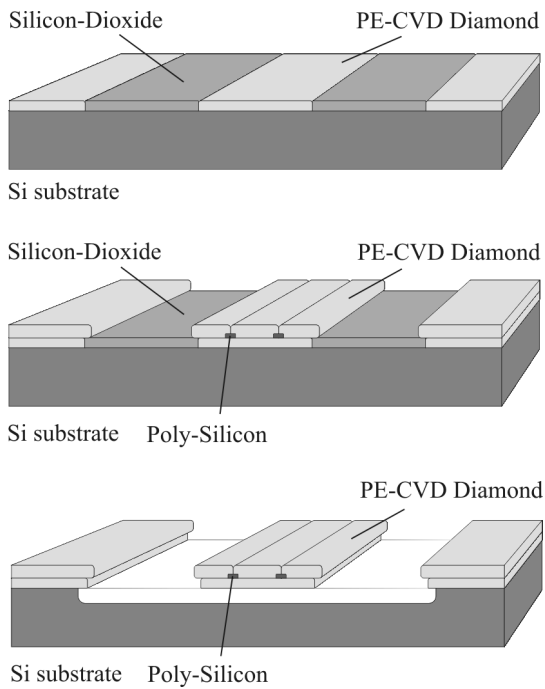

Fig. 2. Process sequence for realisation diamond coated micro-filament heater structure by selective area deposition technique.

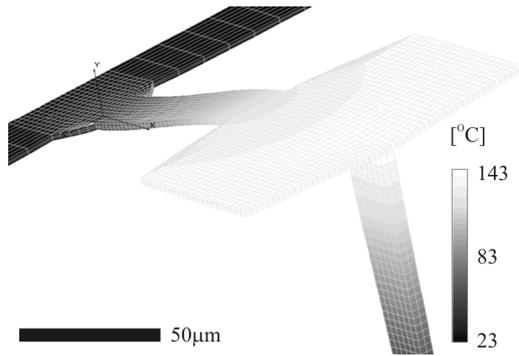

Fig. 4. Simulated temperature distribution of the suspended diamond coated poly-silicon heater filament powered by 100mW.

*Heat conductivity measurement*

The heat conductivity of the structural materials determines the most important thermal characteristics of the structure, the temperature - power consumption function of the filament. [7] By measuring this characteristics the heat conductivity parameters of one structural material can be calculated, assuming the relatively accurate knowledge of the other material parameters and device dimensions. The hotplate temperature can be determined from the resistance of the filament taking into consideration its known temperature coefficient (TC), although this method provides average temperature information. Note, that the TC-T function of the deposited thin layers must be determined first, as it completely differs from available bulk data. Figure 5 represents the temperature vs. power function of the bare heater filaments. The heat loss is developed by heat transfer to the bulk silicon via heat conduction in the suspending beams, by conduction and convection in the surrounding gas and radiation from the hot surfaces.

For extraction of thermal conductivity of silicon-nitride and diamond layers the temperature distribution was also calculated by <u>f</u>inite <u>e</u>lement <u>m</u>odelling (FEM) utilising the ANSYS code (Figure 4.). First order boundary conditions were applied at the suspension/substrate joints by adjusting 23°C, assuming excellent heat conductivity through the silicon substrate. Different heater power levels were set as thermal load for the simulation. The lateral temperature distribution of the heater was calculated by solving the heat conduction equations ignoring heat convection and radiation phenomena.

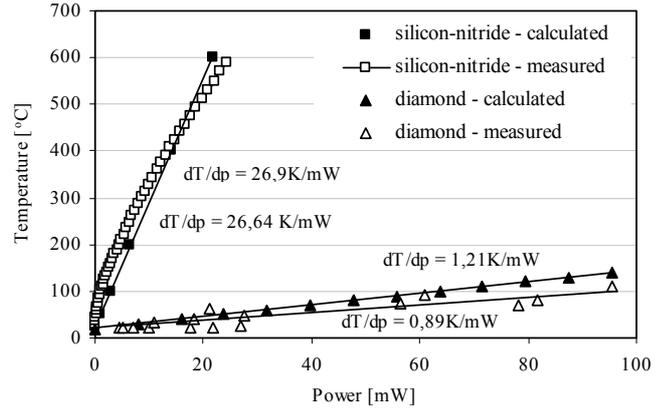

Fig. 5. Temperature vs. power characteristics of the analysed silicon-nitride and diamond coated heater filaments.

In case of FEM simulation only the heat conduction through the suspension beams was taken into consideration. The maximal lateral temperature values simulated are presented in Figure 5 as calculated curves in booth case, applying a preliminary conductivity parameter based on literature data (see Table 1). [8, 9]

The results of the measurements are in accordance with the preliminary calculations. The average specific heat loss of the silicon-nitride and diamond structures are $K_{SiNx}$=0.037mW/K and $K_{diamond}$=1.12mW/K, respectively. The calculated heat conductivity parameters of the silicon-nitride and diamond layers deposited are also presented in Table 1.

*Transient thermal analysis*

The time dependent thermal response of the heated structure is also affected by the physical parameters of the structural materials: as heat conductivity and specific heat, considering the charging and discharging effects of the mass-dependent thermal capacitance and the thermal isolation of the structure. Through the analysis of the dynamic characteristics of the manufactured structures one more parameter (specific heat) can be extracted by determination the response time. In our approach the specific heat of the deposited non-stoichiometric silicon-nitride was determined utilising the thermal equivalent circuit model phenomena. The method of linear RC network theory is also based on the detailed distribution of the analysed physical structures. [10, 11] This tool utilises the analogy of the thermal and electrical conductivity for description of the thermal behaviour of the structure. After definition of thermal resistances and capacitances of the distributed parts based on the material properties, the temperature and thermal flux distribution can be calculated. The temperature of the heater driven by a square wave function can be expressed by the exponential equation:

$$\Delta T(t) = \Delta T_0 e^{-t/RC} + R\Phi_h (1 - e^{-t/RC}) \quad (1)$$

where $\Delta T_0$ is the initial temperature difference between the heater and the ambience, $\Phi_h$ is the driving power pulse (18mW), $R=1/K$ and $C=c*m$ are resultant thermal resistance, and capacitance of the structure. $c$ and $m$ denotes the specific heat and the mass of the structural material, respectively.

For analysing the dynamic thermal response of the filament the heater was driven by a square wave current pulse. The response was recorded using the lock-in thermographic system of Thermosensorik GmbH. [12, 13] The local radiation intensities were detected through several periods by adequately shifting the sampling windows "across the heating-pulse" for obtaining the IR information from the filament in the rising, stable and falling T regime, respectively. According to the predicted time constant, the driver pulse width was set to 10ms to ensure reaching the full temperature saturation, and the heater power was 18mW. For the sake of accuracy the intensity was detected and averaged through several periods. The time dependent response of the filament is represented by the temperature vs. time plot calculated from the emitted IR intensity as unsupported curve in Figure 6.

According to the preliminary calculations, and detailed investigations, a micro filament structure can be properly characterized by a single dominant time constant ($\tau \sim$ 1.11ms), and an RC pair, consequently. The excellent time response – low time constant – can be explained by the outstanding thermal isolation and the low thermal capacitance of the structure. The dynamic properties of the analyzed micro heater structure are also listed in Table 1.

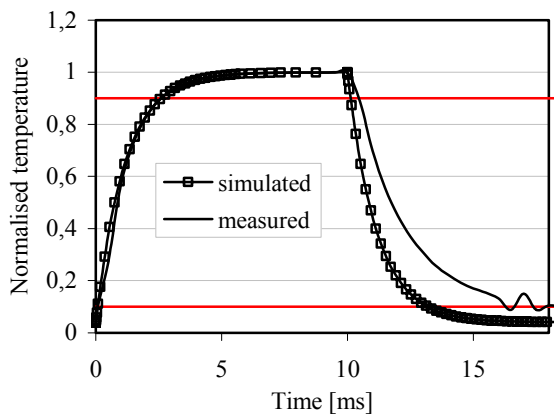

Fig. 6. Measured time dependent temperature response of different heater filaments embedded in silicon-nitride to a 10ms power pulse of 18mW.

A simplified thermal equivalent circuit can be used in the Proteus ISIS Professional 6.1 simulation software of the Labcenter Electronics Corporation to identify the parameters of the additional thermal elements of the structure, like the supporting silicon stab of the silicon-nitride structure described in the introduction. The single RC pair, representing the simplified non-supported structure was supplemented by an additional thermal resistance and capacitance. These are connected parallel according to the thermal influence, the extra heat conduction and capacitance of the thin silicon membrane. First order boundary condition has to be applied for the bulk silicon substrate defined at ground potential. By comparing the two measured temperature response characteristics and time constants both the heat conductivity and the specific heat of the electronically depleted silicon can also be estimated.

TABLE I
CALCULATED THERMAL PROPERTIES OF THE STRUCTURAL MATERIALS APPLIED

| **Silicon-nitride coated filament** (unsupported) | |
|---|---|
| Thermal resistance | 26.9 K/mW |
| **SiN$_x$ thermal conductivity** | **30.825 W/mK** |
| Literature | 9.2 – 30.1 W/mK |
| Time constant | 1.11 ms |
| Thermal capacitance | 41.57 nJ/K |
| **SiN$_x$ specific heat** | **695.76 nJ/kgK** |
| Literature | 710.6 nJ/kgK |
| | |
| **Diamond coated filament** (unsupported) | |
| Thermal resistance | 0.89 K/mW |
| **Diamond thermal conductivity** | **1264 W/mK** |
| Literature | 260 – 895 W/mK |

The preliminary calculations show the limited applicability of the diamond coated filaments in high temperature applications due to its the extreme high thermal conductivity leading to enhanced power consumption of the device.

## V. CONCLUSIONS

Non-stoichiometric silicon-nitride and MWCVD diamond layers were used for reliable processing micro-machined hotplates in order to provide effective chemical and mechanical protection of the MEMS elements.

Alternative structures were applied to determine the thermal parameters of the deposited CVD layers by comparison the results of the thermal measurements and FEM simulations. By fitting the parameters applied for calculations, the static and dynamic behaviour of the model structure was synchronised to the experimental results to extract physical properties of the structural materials.

The proposed method provides uncomplicated and reliable determination of thermal parameters without fabrication of complex test structures. Therefore, it facilitates both structure design and material characterisation by using the functional MEMS devices.

ACKNOWLEDGMENT

This research was partially supported by the OTKA grants T 034821, T 047002 and F 61583.

REFERENCES

[1] Cs. Dücső, M. Ádám, P.Fürjes, M. Hirschfelder, S. Kulinyi and I. Bársony: Explosion-proof monitoring of hydrocarbons by micropellistor, Sensors and Actuators B 95 (2003) pp. 188-193
[2] P. Fürjes, G. Légrádi, Cs. Dücső, A. Aszódi and I. Bársony: Thermal Characterisation of a Direction Dependent Flow Sensor, Sensors and Actuators A 115 (2004) pp. 417-423
[3] P. Fürjes, Cs. Dücső, M. Ádám, A. Morrissey, I. Bársony: „Materials and Processing for Realisation of Micro-hotplates Operated at Elevated Temperature", Journal of Micromechanics and Microengineering 12 (2002) 425-429
[4] Cs. Dücső, et al.: Porous Si bulk micro-machining for thermally isolated membrane formation, Sensors and Actuators A 60 (1997) pp. 235–243
[5] H. Csorbai, P. Fürjes, Gy. Hárs, Cs. Dücső, I. Bársony, E. Kálmán, P. Deák: "Microwave-CVD diamond protective coating for 3D structured silicon microsensors", Materials Science Forum 414-4 (2003) 69-73.
[6] P. Fürjes, Gy. Bognár and I. Bársony: „Powerful tools for thermal characterisation of MEMS", Sensors and Actuators B: Chemical, vol. 120 (1) pp. 270-277, 2006
[7] P. Fürjes, Cs. Dücső, M. Ádám, J. Zettner, I. Bársony: Thermal characterisation of micro-hotplates used in sensor structures, Superlattices and Microstructures 35 (2004) pp. 455-464
[8] E. Jansen, O.Dorsh, E. Obermeier, W. Kulish: „Thermal conductivity measurements on diamond films based on micromechanical devices" Diamond and Related Materials 5 (1996) 644-648
[9] Dean-Mo Liu, Bor-Wen Lin: Thermal Conductivity in Hot-Pressed Silicon Carbide, Ceramics International 22 (1996) pp. 407-414
[10] V. Székely: On the representation of infinite length distributed RC one-ports", IEEE Trans. on Circuits and Systems, Vol. 38 (1991) pp. 711-719
[11] M. Rencz, V. Székely, A. Poppe, G. Farkas, B. Courtois: New Methods and Supporting Tools for the Thermal Transient Testing of Packages, APACK 2001, Singapore, 2001
[12] I. Bársony, P. Fürjes, M. Ádám, Cs. Dücső, J. Zettner and F. Stam: „Thermal response of microfilament heaters in gas sensing", Sensors and Actuators B 103 442-447, 2004
[13] Huth S, Breitenstein O, Huber A, et al.: Lock-in IR-thermography - A novel tool for material and device characterization, Sol. St. Phen. 82-84 (2000) pp. 741-746